# Mass splittings and matrix elements of mesons and baryons containing a single heavy quark

*UKQCD Collaboration* - presented by David Richards[a]

[a]Department of Physics and Astronomy, The University of Edinburgh, Edinburgh EH9 3JZ, Scotland

We present a study in the quenched approximation of the $B$ parameter $B_B$ and the decay constant $f_B$ using heavy-quark propagators implemented in the static approximation, and light-quark propagators computed using an $O(a)$-improved fermion action. We find a value of $B_B^{\rm stat}$ close unity, and discuss the systematic errors entering into the calculation. $f_B^{\rm stat}$ is extracted using a variational fitting technique in order to obtain a reliable estimate of the ground state.

In the second part of the talk, we describe an exploratory study of baryons containing a single heavy quark, computed using the $O(a)$-improved fermion action. We obtain masses generally in good agreement with experiment in both the charm and beauty sectors. We also report preliminary results for the form factor $G_1$ in the semi-leptonic $\Lambda_b \to \Lambda_c$ transition.

## 1. INTRODUCTION

The investigation of hadrons containing a single heavy quark affords the opportunity to determine the least well known elements of the CKM matrix. However, their determination is contingent upon a precise understanding of the effect of the strong interaction on weak decays. This has provided the spur to an intense study of heavy-quark physics in lattice gauge theory. As well as their phenomenological importance, lattice studies of heavy-quark systems may also answer important theoretical questions such as the nature and applicability of the Heavy Quark Effective Theory (HQET). In this talk, we present two related lattice studies of heavy-quark systems. The first is an investigation of the leptonic decays and mixing parameter of the $B$ system, using heavy quarks computed in the static approximation. The second is an investigation of the spectrum of baryons containing a single heavy quark, and the first results of a study of the semi-leptonic form factors of the $\Lambda_b \to \Lambda_c$ transition. Here we use propagating heavy quarks implemented using an $O(a)$-improved fermion action. We conclude this introduction with an outline of details of the lattice simulation common to both studies.

The calculation is performed on 60 configurations of a $24^3 \times 48$ lattice at $\beta = 6.2$. The light-quark propagators are computed using the Sheikholeslami-Wohlert (SW) tree-level $O(a)$-improved fermion action [1], with "rotated" propagators so that the leading discretisation errors are $O(a\alpha_s)$ for light-quark systems [2]. Propagators are computed at three values of the light-quark mass, corresponding to $\kappa_l = 0.14144$, 0.14226 and 0.14262. These lie in the region of the strange quark, whose mass as determined from the ratio $m_K^2/m_\rho^2$ corresponds to $\kappa_s = 0.1419(1)$; massless quarks correspond to $\kappa_{\rm crit} = 0.14315(2)$ [3].

For dimensionful quantities, we take the inverse lattice spacing to be $a^{-1} = 2.9(2)\,{\rm GeV}$, where the quoted error is sufficient to encompass the values determined from $f_\pi$, $R_0$, $m_\rho$ and the string tension.

## 2. $B_B$ AND $f_B^{\rm stat}$

The static approximation arises from keeping only the leading term in the expansion of the heavy-quark propagator [4]. In this calculation, the static quark propagators are computed from "smeared", or spatially-extended, sources to both local and smeared sinks; smearing has proved essential to obtain a signal in this case [5]. The light-quark propagators are computed using local sources and sinks. We use both gauge-invariant

Jacobi smearing [6], with an r.m.s smearing radius of $r_0 \sim 6.4$, and several Coulomb-gauge-fixed smearing functions.

### 2.1. Lattice operators and renormalisation

In the continuum full theory, the pseudoscalar decay constant of the $B$ meson is defined through

$$\langle 0|A_\mu(0)|B(\vec{p})\rangle = ip_\mu f_B \quad (1)$$

while the $B$-parameter is defined through

$$B_B(\mu) = \frac{\langle \bar{B^0}|O_L(0)|B^0\rangle_\mu}{\frac{8}{3}f_B^2 M_B^2} \quad (2)$$

where $O_L$ is the $\Delta B = 2$ four-fermi operator

$$O_L = \left(\bar{b}\gamma_\mu(1-\gamma_5)q\right)\left(\bar{b}\gamma_\mu(1-\gamma_5)q\right), \quad (3)$$

and $\mu$ is the renormalisation scale. The renormalisation-group-invariant (RGI) $B$ parameter is then defined by

$$B_B = \alpha_s(\mu)^{-2/\beta_0} B_B(\mu). \quad (4)$$

The matching of these operators in the full continuum theory to the relevant operators in the static lattice theory is performed as a two stage process: the matching of operators in the full continuum theory to operators in the static continuum theory, followed by the matching of operators in the static continuum theory to those in the static lattice theory. Here we will illustrate the procedure in the case of the $B$ parameter.

The one-loop matching factor between the continuum full theory at a scale $m_b$, and the continuum effective theory at a scale $\mu < m_b$ introduces mixing with the operator $O_S \equiv \left(\bar{b}(1-\gamma_5)q\right)\left(\bar{b}(1-\gamma_5)q\right)$, and assumes the form

$$O_L^{\rm full}(m_b) = \frac{g^2}{16\pi^2}C_S O_S^{\rm eff}(\mu) + \left\{1 + \frac{g^2}{16\pi^2}\left[4\ln(m_b^2/\mu^2) + C_L\right]\right\} O_L^{\rm eff}(\mu) \quad (5)$$

with $C_S = -8$ and $C_L = -14$ [7].

The matching of the static continuum theory to the static lattice theory introduces mixing with two further operators through the breaking of chiral symmetry

$$O_L^{\rm eff} = \left\{1 + \frac{\alpha_s^{\rm latt}(a^{-1})}{4\pi}D_L\right\} O_L^{\rm latt} + \frac{\alpha_s^{\rm latt}(a^{-1})}{4\pi}D_R O_R^{\rm latt} + \frac{\alpha_s^{\rm latt}(a^{-1})}{4\pi}D_N O_N^{\rm latt} \quad (6)$$

where $D_L, D_R$ and $D_N$ are listed in ref. [7] and [8]; note that $D_L$ is corrected to use the reduced value of the quark self-energy [9]. Expanding Equation (5) and (6) to $O(\alpha_s)$, and using the boosted value of $\alpha_s^{\rm latt}(a^{-1})$ [10], we obtain

$$O_L^{\rm full}(m_b) = \sum_{i=L,N,R,S} Z_i O_i^{\rm latt} \quad (7)$$

with $Z_L = 0.55$, $Z_N = -0.15$, $Z_R = -0.04$ and $Z_S = -0.18$.

We note that the $O(\alpha_s)$ contributions to the matching coefficient are, in general, large, questioning the applicability of the one-loop perturbation theory calculation. Indeed, if we first compute $O_L^{\rm eff}(a^{-1})$ according to Equation (6), and insert the result into Equation (5), we obtain a result for $B_B$ some 20 % larger than that quoted below, thus providing an estimate of the possible uncertainty due to higher order perturbative corrections.

The calculation of $Z_A^{\rm stat}$, the matching factor for $f_B^{\rm stat}$, follows the procedure described above, with the simplification that there is no operator mixing. Here we use $Z_A^{\rm stat} = 0.78$, very close to the values employed in our previous study [11] and by APE [12].

### 2.2. Calculation of $f_B^{\rm stat}$

To extract $f_B^{\rm stat}$, we study the two-point correlators:

$$\begin{aligned} C^{SS}(t) &= \sum_{\vec{x}} \langle 0|A_4^S(\vec{x},t)A_4^{S\dagger}(\vec{0},0)|0\rangle \\ &\xrightarrow{t\gg 0} (Z^S)^2 e^{-Et} \end{aligned} \quad (8)$$

$$\begin{aligned} C^{LS}(t) &= \sum_{\vec{x}} \langle 0|A_4^L(\vec{x},t)A_4^{S\dagger}(\vec{0},0)|0\rangle \\ &\xrightarrow{t\gg 0} Z^S Z^L e^{-Et} \end{aligned} \quad (9)$$

where $E$ is the (unphysical) binding energy, and $S$ and $L$ refer to operators computed using smeared and local heavy-quark fields respectively. $Z^L$ is computed from the ratio of $C^{LS}$ and $C^{SS}$, with $E$ and $Z^S$ extracted from a fit to $C^{SS}$. The pseudoscalar decay constant $f_B^{\rm stat}$ is then related to $Z^L$ via

$$f_B^{\rm stat} = Z^L\sqrt{\frac{2}{M_B}}Z_A^{\rm stat}. \quad (10)$$

In order to ensure that we have isolated the ground state reliably, for the case of gauge-fixed smearing we compare the values for $Z_L$ obtained from a singe-state fit, to those obtained to a two-state fit to a matrix of correlators [13]. We find consistency both between one-state and two-state fits, and between fits with differing smearing functions. We quote as our best estimate the value obtained from a two-state fit to the exponentially-smeared gauge-fixed correlators, where the statistical errors are sufficient to encompass the values from the different fitting procedures and smearing functions. We obtain $Z^L = 0.112\ {}^{+8}_{-8}$, yielding

$$f_B^{\rm stat} = 266\ {}^{+18}_{-20}\,({\rm stat})\ \ {}^{+28}_{-27}\,({\rm syst}); \qquad (11)$$

$$\frac{f_{B_s}}{f_{B_d}} = 1.16\ {}^{+4}_{-3}, \qquad (12)$$

where the second error on $f_B^{\rm stat}$ arises from the uncertainty on the scale. The salient feature is the agreement between $Z_L$ obtained from simulations in which the light quarks are computed using the Wilson and using the SW action [12]; the discrepancy between the central values of $f_B^{\rm stat}$ obtained with the two actions arises principally from the very different perturbative values of $Z_A^{\rm stat}$.

### 2.3. $B$ parameter $B_B$

We compute the $B$ parameter by evaluating the three-point correlators

$$K_i^{SS}(t_1,t_2) \equiv \sum_{\vec{x_1},\vec{x_2}} \langle 0|T{A^\dagger}^S(\vec{x}_1,-t_1)\,O_i^{\rm latt}(0)\,{A^\dagger}^S(\vec{x}_2,t_2)|0\rangle$$

$$\stackrel{t_1,t_2\gg 0}{\longrightarrow} \frac{(Z^S)^2}{2\,M_B}\,{\rm e}^{-E\,(t_1+t_2)}\langle\overline{B^0}|O_i^{\rm latt}|B^0\rangle \qquad (13)$$

where $i = L, R, S, N$. The amplitude $Z_S$ and the exponential factors are eliminated by studying the ratios

$$R_i^{SS}(t_1,t_2) = \frac{K_i^{SS}(t_1,t_2)}{\frac{8}{3}\,C^{SL}(t_1)\,C^{SL}(t_2)} \qquad (14)$$

$$\stackrel{t_1,t2\gg 0}{\longrightarrow} \quad R_i \equiv \frac{\langle\bar{B}^0|O_i^{\rm latt}|B^0\rangle}{\frac{8}{3}(Z^L)^2}. \qquad (15)$$

We then have

$$B_B^{\rm stat}(m_b) = \sum_i Z_i (Z_A^{\rm stat})^{-1} R_i \qquad (16)$$

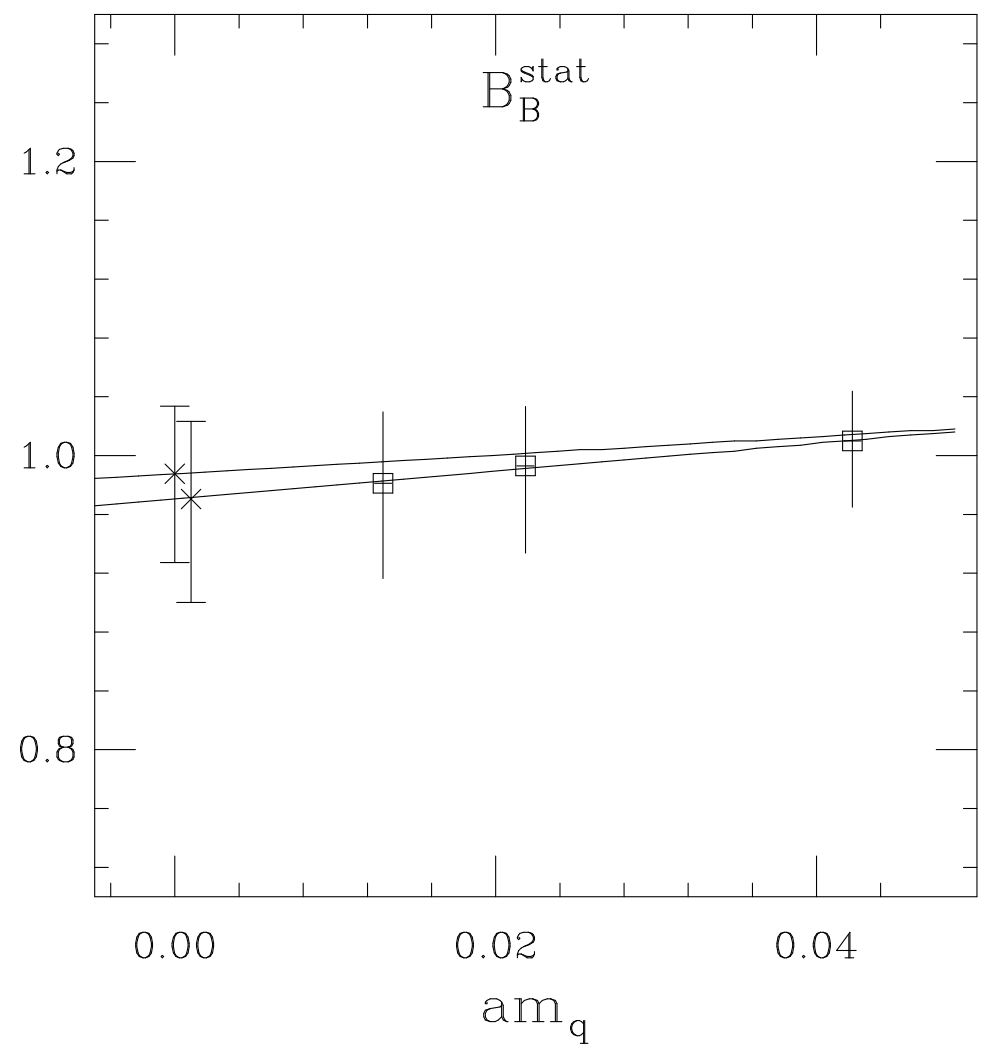

Figure 1. The $B$ parameter at each $\kappa_l$ is shown as the squares for gauge-invariant Jacobi smearing. The solid (dashed) lines show the correlated (uncorrelated) chiral extrapolations respectively, and the crosses the extrapolated values.

where the $Z_i$ are given following Equation (7).

The RGI $B$ parameter at each $\kappa_l$, together with the chiral extrapolations, are shown in Figure 1 for the gauge-invariant smearing procedure. This procedure, with a correlated chiral extrapolation, provides our best value for $B_B^{\rm stat}$:

$$B_{B_d} = 0.99\ {}^{+5}_{-6}\,({\rm stat})\ \ {}^{+3}_{-2}\,({\rm syst}) \qquad (17)$$

$$B_{B_s} = 1.01\ {}^{+4}_{-5}\,({\rm stat})\ \ {}^{+2}_{-1}\,({\rm syst}) \qquad (18)$$

where the systematic errors encompass the variations due to different smearing functions, different fitting techniques and correlated or uncorrelated chiral extrapolations. However, it must be emphasised that these values increase by 20 % if the alternative matching prescription discussed in Section 2.1 is adopted, and this must be regarded as a further systematic uncertainty in our calculation.

## 3. HEAVY BARYON PHYSICS

For the study of baryons containing a single heavy quark, we employ propagating quarks com-puted using the SW action, at four values of the heavy-quark mass around that of the charm quark, corresponding to $\kappa_h = 0.121$, 0.125, 0.129 and 0.133. Heavy quark propagators are computed from Jacobi-smeared sources to both local and smeared sinks; in contrast to the preceding study, we also employ light quark propagators from smeared sources, but only at our two heaviest light-quark masses.

### 3.1. Lattice operators

There are eight baryons containing one heavy and two light quarks, and these are listed in Table 1, together with their quantum numbers and quark content. At lowest order in HQET it is possible to distinguish the quantum numbers of the light degrees of freedom from those of the heavy quark, and these also are shown in the table.

The spectrum can be computed using the following operators:

$$\mathcal{O}_5 = \epsilon_{abc}(l^a \mathcal{C}\gamma_5 l'^b)h^c \quad (19)$$

$$\mathcal{O}_\mu = \epsilon_{abc}(l^a \mathcal{C}\gamma_\mu l^b)h^c \quad (20)$$

$$\mathcal{O}'_\mu = \epsilon_{abc}(l^a \mathcal{C}\gamma_\mu l'^b)h^c, \quad (21)$$

where $\mathcal{C}$ is the charge conjugation matrix, $l$, $l'$ are light-quark fields, and $h$ is the heavy-quark field.

The operator $\mathcal{O}_5$ corresponds to $s_l^{\pi_l} = 0^+$ spin-parity for the light degrees of freedom and a total spin-parity for the baryon $J^P = \frac{1}{2}^+$. We extract the mass from a fit to the two-point correlator

$$G_5(\vec{p},t) = \sum_x \mathrm{e}^{-\vec{p}\cdot\vec{x}} \langle \mathcal{O}_5(\vec{x},t)\bar{\mathcal{O}}_5(\vec{0},0)\rangle \quad (22)$$

$$\xrightarrow{t\gg 0} \frac{Z^2(|\vec{p}|)}{2E}\mathrm{e}^{-Et}(\not{p}+M). \quad (23)$$

The operator $\mathcal{O}_\mu^{(\prime)}$ creates both spin $\frac{3}{2}$ and spin $\frac{1}{2}$ particles. However, projectors $P^{3/2}$ and $P^{1/2}$ can be defined [14] to separate the spin $\frac{3}{2}$ and spin $\frac{1}{2}$ contributions to the correlator of $\mathcal{O}_\mu$,

$$G_{\mu\nu}(\vec{p},t) = \sum_{\vec{x}} \mathrm{e}^{-i\vec{p}\cdot\vec{x}} \langle \mathcal{O}_\mu(\vec{x},t)\bar{\mathcal{O}}_\nu(\vec{0},0)\rangle \quad (24)$$

$$\xrightarrow{t\gg 0} \frac{Z_{\frac{3}{2}}}{E_{\frac{3}{2}}}\mathrm{e}^{-E_{\frac{3}{2}}t}(\not{p}+M_{\frac{3}{2}})P^{\frac{3}{2}}_{\mu\nu} + \frac{Z_{\frac{1}{2}}}{E_{\frac{1}{2}}}\mathrm{e}^{-E_{\frac{1}{2}}t}(\not{p}+M_{\frac{1}{2}})P^{\frac{1}{2}}_{\mu\nu}. \quad (25)$$

### 3.2. Masses and mass splittings

The masses are obtained from single-state fits to both the LS and SS correlators of Equations (22) and (24). In general, we find consistency between the fits, but those to the SS correlators are generally more stable under variations in the fitting range, and have lower $\chi^2$/dof. We therefore use these for our best values.

The heavy-baryon masses at fixed $\kappa_h$ are extrapolated linearly in the sum of the light-quark masses to the chiral limit; because we have correlators for both degenerate and non-degenerate light-quark masses, we can include three points in the extrapolation.

Following HQET, the chirally-extrapolated masses are extrapolated linearly in the heavy-light pseudoscalar mass to obtain the various baryon masses at the physical $D$ and $B$ masses; the extrapolation proved insensitive to different modelling functions. The masses obtained from this procedure, together with the experimental values where known, are shown in Table 1.

To focus on the dynamics of the light quarks and their interaction with the heavy quark, it is useful to study the mass splittings. Both the $\Lambda - \mathrm{PS}$ and the $\Sigma - \Lambda$ mass splittings are sensitive to the dynamics of the light degrees of freedom. Furthermore, in the latter case we expect no dependence on the heavy-quark mass at $O(1/m_h)$. Below we give the dimensionless ratios for these splittings at both the charm and the beauty masses:

| | latt. | Exp. |
|---|---|---|
| $(\Lambda_c - D)/(\Lambda_c + D)$ | $0.099^{+1}_{-2}$ | 0.100(3) |
| $(\Lambda_b - B)/(\Lambda_b + B)$ | $0.033^{+1}_{-2}$ | 0.033(3) |
| $(\Sigma_c - \Lambda_c)/(\Sigma_c + \Lambda_c)$ | $0.039^{+1}_{-2}$ | 0.035(3) |
| $(\Sigma_b - \Lambda_b)/(\Sigma_b + \Lambda_b)$ | $0.017^{+1}_{-2}$ | 0.016(2) |

We have little confidence in our measurements of the spin splittings; they have large relative statistical errors, and it is difficult to perform the extrapolation in the heavy-quark mass. Nevertheless, our measurements of the spin splittings are generally inconsistent with experiment, and in most cases consistent with zero, mirroring the well-known feature in the meson system.

| | $(S)$ | $J^P$ | $(I)$ | $s_l^{\pi_l}$ | Quarks | Op. | $h$=charm Exp. | $h$=charm Latt. | $h$ = beauty Exp. | $h$ = beauty Latt. |
|---|---|---|---|---|---|---|---|---|---|---|
| $\Lambda_h$ | $(0)$ | $\frac{1}{2}^+$ | $(0)$ | $0^+$ | $(ud)h$ | $\mathcal{O}_5$ | 2.285(1) | $2.28\ ^{+4}_{-4}\ ^{+3}_{-3}$ | 5.64(5) | $5.59\ ^{+9}_{-10}\ ^{+3}_{-3}$ |
| $\Sigma_h$ | $(0)$ | $\frac{1}{2}^+$ | $(1)$ | $1^+$ | $(uu)h$ | $\mathcal{O}_\mu$ | 2.453(1) | $2.45\ ^{+6}_{-4}\ ^{+5}_{-5}$ | 5.81(6) | $5.69\ ^{+10}_{-10}\ ^{+5}_{-4}$ |
| $\Sigma_h^*$ | $(0)$ | $\frac{3}{2}^+$ | $(1)$ | $1^+$ | $(uu)h$ | $\mathcal{O}_\mu$ | 2.530(7) | $2.43\ ^{+5}_{-4}\ ^{+5}_{-4}$ | 5.87(6) | $5.68\ ^{+9}_{-10}\ ^{+5}_{-4}$ |
| $\Xi_h$ | $(-1)$ | $\frac{1}{2}^+$ | $(\frac{1}{2})$ | $0^+$ | $(us)h$ | $\mathcal{O}_5$ | 2.468(4) | $2.41\ ^{+3}_{-3}\ ^{+4}_{-5}$ | | $5.69\ ^{+7}_{-10}\ ^{+4}_{-4}$ |
| $\Xi_h'$ | $(-1)$ | $\frac{1}{2}^+$ | $(\frac{1}{2})$ | $1^+$ | $(us)h$ | $\mathcal{O}_\mu'$ | 2.560 | $2.58\ ^{+5}_{-4}\ ^{+5}_{-6}$ | | $5.85\ ^{+10}_{-12}\ ^{+5}_{-5}$ |
| $\Xi_h^*$ | $(-1)$ | $\frac{3}{2}^+$ | $(\frac{1}{2})$ | $1^+$ | $(us)h$ | $\mathcal{O}_\mu'$ | 2.643(2) | $2.55\ ^{+5}_{-4}\ ^{+5}_{-5}$ | | $5.82\ ^{+8}_{-11}\ ^{+5}_{-5}$ |
| $\Omega_h$ | $(-2)$ | $\frac{1}{2}^+$ | $(0)$ | $1^+$ | $(ss)h$ | $\mathcal{O}_\mu$ | 2.704(20) | $2.68\ ^{+4}_{-4}\ ^{+6}_{-5}$ | | $5.69\ ^{+10}_{-10}\ ^{+6}_{-5}$ |
| $\Omega_h^*$ | $(-2)$ | $\frac{3}{2}^+$ | $(0)$ | $1^+$ | $(ss)h$ | $\mathcal{O}_\mu$ | | $2.64\ ^{+4}_{-3}\ ^{+6}_{-6}$ | | $5.89\ ^{+6}_{-8}\ ^{+6}_{-6}$ |

Table 1
Quantum numbers of the eight baryons containing a single heavy quark $h$. $I$, $s_l^{\pi_l}$ are the isospin and the spin-parity of the light degrees of freedom and $S$, $J^P$ are the total strangeness and the spin-parity. The lattice detemination is given in GeV, together with the experimental values, where known.

### 3.3. Semi-leptonic form factor

The matrix elements for the decay $\Lambda_b \to \Lambda_c l \bar{\nu}_l$ are described by six form factors:

$$\langle \Lambda_h^s(\vec{p}) | J_\mu^{h,h'}(y) | \Lambda_{h'}^{s'}(\vec{p}') \rangle = \bar{h}_h^{(s)}(\vec{p}) \mathcal{F}_\mu^{h,h'}(p',p) u_{h'}^{(s')}(\vec{p}') \quad (26)$$

where

$$J_\mu^{h,h'}(x) = \bar{h}(x)\gamma_\mu(1-\gamma_5)h'(x) \quad (27)$$

and

$$\mathcal{F}_\mu^{h,h'}(p',p) = \left(F_1(\omega)\gamma_\mu + F_2(\omega)v'_\mu + F_3(\omega)v_\mu\right) - \left(G_1(\omega)\gamma_\mu + G_2(\omega)v'_\mu + G_3(\omega)v_\mu\right)\gamma_5. \quad (28)$$

Here $s^{(')}$ are helicity indices, $v^{(')}$ are the four-velocities, $u$'s are the heavy-baryon spinors and $\omega = v.v'$. $F_i(\omega)$ and $G_i(\omega)$ are form factors of the vector and axial-vector currents respectively. HQET requires that the form factors be related to a universal RGI function $\xi(\omega)$, which we call the baryon Isgur-Wise function:

$$F_i(\omega) = (\alpha_i + \beta_i(\omega) + \gamma_i(\omega))\,\xi(\omega) \quad (29)$$

$$G_i(\omega) = \left(\alpha_i^5 + \beta_i^5(\omega) + \gamma_i^5(\omega)\right)\xi(\omega) \quad (30)$$

$$\alpha_1 = \alpha_1^5 = 1; \alpha_{2,3} = \alpha_{2,3}^5 = 0. \quad (31)$$

$\beta_i^{(5)}(\omega)$ parametrise the radiative corrections, whilst $\gamma_i^{(5)}(\omega)$ contain corrections proportional to powers of the inverse heavy-quark mass.

The form factor $G_1(\omega)$ has no $1/m_h$ corrections at $\omega = 1$. Furthermore, this form factor dominates the decay rate near zero recoil, affording the possibility of an accurate experimental determination of $V_{cb}$. In order to compute $G_1(\omega)$, we evaluate the three-point correlator

$$C_\mu(t_x,t_y) = \sum_{\vec{x},\vec{y}} \mathrm{e}^{-i\vec{p}\cdot\vec{x}} \mathrm{e}^{-i\vec{q}\cdot\vec{y}} \langle \mathcal{O}_5^h(x) J_\mu^{h,h'}(y) \bar{\mathcal{O}}_5^{h'}(0) \rangle. \quad (32)$$

By taking the ratio of this to an appropriate product of two-point functions we obtain $\mathcal{F}_\mu^{h,h'}(p',p)^{\mathrm{latt}}$, the lattice analogue of $\mathcal{F}_\mu^{h,h'}(p',p)$ with $\vec{p}' = \vec{p} + \vec{q}$, and hence extract $G_1^{\mathrm{latt}}(\omega)$.

In the case of degenerate transitions, in which the heavy quarks are of equal mass, we can reduce discretisation errors in $G_1(\omega)$, as well as remove the dependence on the matching coefficient $Z_A$, by dividing $G_1^{\mathrm{latt}}(\omega)$ by the measured $G_1^{\mathrm{latt}}(1)$; the latter is obtained from the transition $(0,0,0) \to (0,0,0)$. In Figure 2, we show preliminary results for $G_1(\omega)$, normalised to $G_1(1) = 1$, at our four value of $\kappa_h$, with both light quarks corresponding to $\kappa_l = 0.14144$. The error bars at zero recoil arise from the measurement of the transition $(\pi/12\,a^{-1},0,0) \to (\pi/12\,a^{-1},0,0)$.

Radiative corrections are not included in this preliminary analysis, so some caution has to be exercised before calling this an "Isgur-Wise" function. Furthermore the power correction $\gamma_i$ are not yet determined in a model-independent way. Nevertheless, the points appear to lie on a universal mass-independent curve, suggesting that mass-dependent corrections are small; this seems reasonable, since $G_1$ is free of $O(1/m_h)$ corrections at zero recoil.

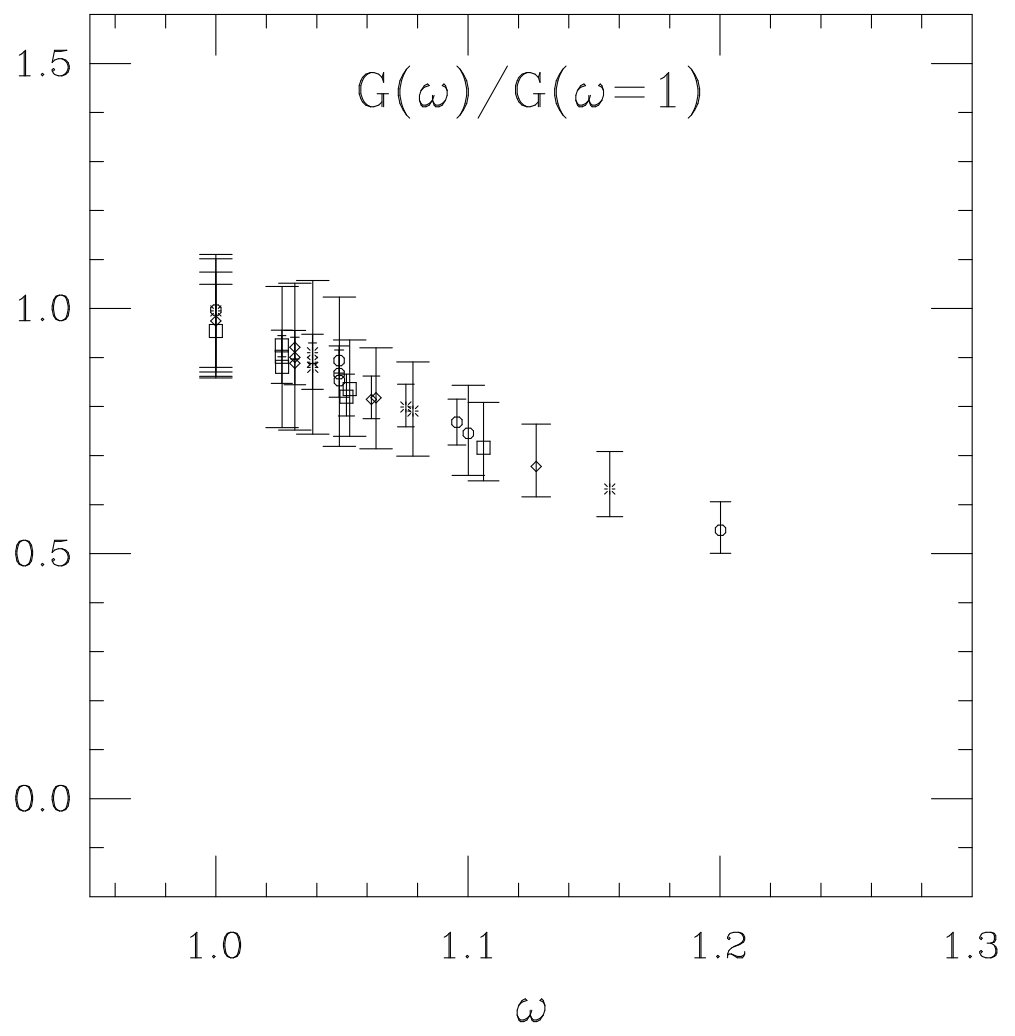

Figure 2. The form factor $G_1^{\rm latt}(\omega)/G_1^{\rm latt}(1)$ for degenerate transitions, at $\kappa_l = 0.14144$. The different plotting symbols correspond to different $\kappa_h$

## 4. CONCLUSIONS

The study of heavy-quark systems on the lattice is vital both as a phenomenological tool to enable the extraction of CKM matrix elements, and as a laboratory in which to explore the heavy-quark symmetry. In this talk, two different studies have been presented. The measurements of $f_B^{\rm stat}$ and $B_B^{\rm stat}$ are remarkably consistent between different interpolating operators and between different fitting techniques. The principle requirement to improve our determination still further is a better understanding of the matching of operators between the continuum and lattice.

The exploratory study of the spectrum of heavy baryons has produced notable agreement with experiment, both in the charm and beauty sectors. Furthermore, we have demonstrated the feasibility of extracting heavy-baryon semi-leptonic form factors, opening the possibility of further independent measurements of $V_{cb}$.

## ACKNOWLEDGEMENTS

DGR acknowledges the support of PPARC through grant GR/J21347 and an Advanced Fellowship. The study of heavy baryons was performed on the 320-node T3D at the University of Edinburgh, with the support of EPSRC grant GR/K41663.